\documentclass[12pt]{article}
\usepackage{amssymb,amsmath,epsfig}

\begin{document}

\title{\bf Energy Conditions in a Generalized Second-Order Scalar-Tensor Gravity}

\author{M. Sharif \thanks{msharif.math@pu.edu.pk} and Saira Waheed
\thanks{smathematics@hotmail.com}\\
Department of Mathematics, University of the Punjab,\\
Quaid-e-Azam Campus, Lahore-54590, Pakistan.}

\date{}

\maketitle
\begin{abstract}
The study of energy conditions has many significant applications in
general relativistic and cosmological contexts. This paper explores
the energy conditions in the framework of the most general
scalar-tensor theory with field equations involving second-order
derivatives. For this purpose, we use flat FRW universe model with
perfect fluid matter contents. By taking power law ansatz for scalar
field, we discuss the strong, weak, null and dominant energy
conditions in terms of deceleration, jerk and snap parameters. Some
particular cases of this theory like k-essence model and modified
gravity theories etc. are analyzed with the help of the derived
energy conditions and the possible constraints on the free
parameters of the presented models are determined.
\end{abstract} {\bf
Keywords:} General scalar-tensor theory; Scalar field; Energy conditions.

\section{Introduction}

``The increasing rate of cosmic expansion in current phase" is one
of the primal facts in modern cosmology that is supported by some
sort of energy with negative pressure as well as hidden
characteristics refereed to dark energy (DE) (its existence is
affirmed by the recent data of many astronomical observations)
\cite{1}. The investigation of this hidden unusual nature of DE has
been carried out in two ways: one approach utilizes the modified
matter sources, that is, different models like Chaplygin gas
\cite{1*}, quintessence \cite{2}, k-essence \cite{3}, cosmological
constant \cite{4} etc. are introduced in the usual matter contents
within the gravitational framework of general relativity (GR) while
in second approach, GR framework is modified by the inclusion of
some extra degrees of freedom \cite{5}. Examples of some well-known
modified gravity theories include $f(R)$ gravity \cite{6},
scalar-tensor theories like Brans-Dicke (BD) gravity \cite{7},
Gauss-Bonnet gravity \cite{8}, $f(T)$ theory \cite{9}, $f(R,T)$
gravity \cite{10} etc. Modified matter sources are rather
interesting but each faces some difficulties and hence could not
prove to be very promising. Modified gravitational theories being
large-distance modifications of gravity have brought a fresh insight
in modern cosmology. Among these, scalar-tensor theories are
considered to be admirable efforts for the investigation of DE
characteristics, which are obtained by adding an extra scalar degree
of freedom in Einstein-Hilbert action.

Scalar field provides a basis for many standard inflationary models,
leading to an effective candidate of DE. In literature \cite{11},
many inflationary models have been constructed like chaotic
inflation, small/large inflation, k-inflation, Dirac-Born-Infeld
(DBI) inflation, single field k-inflation etc. All of these are some
peculiar extensions of the k-essence models. Although the k-essence
scalar models are considered to be the general scalar field theories
described by the Lagrangian in terms of first-order scalar field
derivatives, i.e., $L=L(\phi,\nabla\phi)$. However, Lagrangian with
higher-order scalar field derivatives
$(L=L(\phi,\nabla\phi,\nabla\nabla\phi)$ can be taken into account
which fixes the equations of motion (obtained by metric and scalar
field variations of the Lagrangian density) to second-order
\cite{12,13}. Horndeski \cite{12} was the pioneer to discuss the
concept of most general Lagrangian with single scalar field.
Recently, this action is discussed by introducing a covariant
Galilean field with second-order equations of motion \cite{13}.
Kobayashi et al. \cite{14} developed a correspondence between these
Lagrangians. This theory has fascinated many researchers and much
work has been done in this context, e.g., \cite{14,15}.

The energy conditions have many significant theoretical applications
like Hawking Penrose singularity conjecture is based on the strong
energy condition \cite{16} while the dominant energy condition is
useful in the proof of positive mass theorem \cite{17}. Furthermore,
null energy condition is a basic ingredient in the derivation of
second law of black hole thermodynamics \cite{18}. On the
cosmological grounds, Visser \cite{19} discussed various
cosmological terms like distance modulus, look back time,
deceleration and statefinder parameters in terms of red shift using
energy condition constraints. These conditions are originally
formulated in the context of GR and then extended to modified
theories of gravity. Many authors have explored these energy
conditions in the framework of modified gravity and found
interesting results \cite{20,21*}.

Basically, modified gravity theories contain some extra functions
like higher order derivatives of curvature term or some function of
Einstein tensor or scalar field etc. Thus it is a point of debate
that how one can constrain the added extra degree of freedom
consistently with the recent observations. The energy conditions can
be used to put some constraints on these functions that could be
consistent with those already found in the cosmological arena.
Recently, these energy conditions have been discussed in $f(T)$
\cite{21} and $f(R,T)$ \cite{22} theories.

In this paper, we study the energy condition bounds in a most
general scalar-tensor theory. The paper is designed in the following
layout. Next section defines the energy conditions in GR as well as
in a general modified gravitational framework. Section \textbf{3}
provides basic formulation of the most general scalar-tensor theory.
In the same section, we formulate the energy conditions in terms of
some cosmological parameters within such modified framework. In
section \textbf{4}, we provide some specific cases of this theory
and discuss the corresponding constraints. Finally, we summarize and
present some general remarks.

\section{Energy Conditions}

In this section, we discuss the energy conditions in GR framework
and then express the respective conditions in a general modified
gravity. In GR, the energy conditions come from a well-known purely
geometric relationship known as Raychaudhari equation \cite{18,23}
together with the lineament of gravitational attractiveness. In a
spacetime manifold with vector fields $u^\mu$ and  $k^\mu$ as
tangent vectors to timelike and nulllike geodesics of the
congruence, the temporal variation of expansion for the respective
curves is described by the Raychaudhari equation as
\begin{eqnarray}\label{1}
\frac{d\theta}{d\tau}=-\frac{1}{3}\theta^2-\sigma_{\mu\nu}\sigma^{\mu\nu}
+\omega_{\mu\nu}\omega^{\mu\nu}-R_{\mu\nu}u^\mu u^\nu,
\end{eqnarray}
\begin{eqnarray}\label{2}
\frac{d\theta}{d\tau}=-\frac{1}{3}\theta^2-\sigma_{\mu\nu}\sigma^{\mu\nu}
+\omega_{\mu\nu}\omega^{\mu\nu}-R_{\mu\nu}k^\mu k^\nu.
\end{eqnarray}
Here $~R_{\mu\nu},~\theta,~\sigma^{\mu\nu}$ and $\omega^{\mu\nu}$
represent the Ricci tensor, expansion, shear and rotation,
respectively related with the congruence of timelike or nulllike
geodesics.

The characteristic of the gravity that it is attractive, leads to
the condition $\frac{d\theta}{d\tau}<0$. For infinitesimal
distortions and vanishing shear tensor $\omega_{\mu\nu}=0$, i.e.,
zero rotation (for any hypersurface of orthogonal congruence), we
ignore the second-order terms in Raychaudhari equation and
consequently integration leads to $\theta=-\tau R_{\mu\nu}u^\mu
u^\nu=-\tau R_{\mu\nu}k^\mu k^\nu$. It further implies that
\begin{equation*}
R_{\mu\nu}u^\mu u^\nu\geq0,\quad R_{\mu\nu}k^\mu k^\nu\geq0.
\end{equation*}
Since GR and its modifications lead to a relationship of the matter
contents, i.e., energy-momentum tensor in terms of Ricci tensor
through the field equations, therefore the respective physical
conditions on the energy-momentum tensor can be determined as
follows
\begin{eqnarray}\label{1*}
&&R_{\mu\nu}u^\mu u^\nu=(T_{\mu\nu}-\frac{T}{2}g_{\mu\nu})u^\mu
u^\nu\geq0,
\\\label{2*} &&R_{\mu\nu}k^\mu
k^\nu=(T_{\mu\nu}-\frac{T}{2}g_{\mu\nu})k^\mu k^\nu\geq0,
\end{eqnarray}
where $T_{\mu\nu}$ and $T$ are the energy-momentum tensor and its
trace, respectively. For perfect fluid with density $\rho$ and
pressure $p$ defined by
\begin{eqnarray}\label{3}
T_{\mu\nu}=(\rho+p)u_{\mu}u_{\nu}-pg_{\mu\nu},
\end{eqnarray}
the strong and null energy conditions, respectively, are defined by
the inequalities $\rho+3p\geq0$ and $\rho+p\geq0$, while the weak
and dominant energy conditions are defined, respectively, by
$\rho\geq0$ and $\rho\pm p\geq0$.

Raychaudhari equation being a geometrical statement works for all
gravitational theories. Therefore, its interesting features like
focussing of geodesic congruences as well as the attractiveness of
gravity can be used to derive the energy constraints in the context
of modified gravity. In case of modified gravity, we assume that the
total matter contents of the universe act like perfect fluid and
consequently these conditions can be defined in terms of effective
energy density and pressure (matter sources get modified and we
replace $T_{\mu\nu}$ and $T$ in Eqs.(\ref{1*}) and (\ref{2*}) by
$T^{eff}_{\mu\nu}$ and $T^{eff}$, respectively). These conditions
can be regarded as an extension of the respective conditions in GR
given by \cite{21*}
\begin{eqnarray}\nonumber
\textbf{NEC}:\quad&&\rho^{eff}+p^{eff}\geq0,\\\nonumber
\textbf{SEC}:\quad&&\rho^{eff}+p^{eff}\geq0,\quad\rho^{eff}+3p^{eff}\geq0,\\\nonumber
\textbf{WEC}:\quad&&\rho^{eff}\geq0,\quad\rho^{eff}+p^{eff}\geq0,\\\label{4}
\textbf{DEC}:\quad&&\rho^{eff}\geq0,\quad\rho^{eff}\pm
p^{eff}\geq0.
\end{eqnarray}
For a detailed discussion, we suggest the readers to study a recent
paper \cite{23*}.

The DE requires negative EoS parameter $\omega\leq-1/3$, for the
explanation of cosmic expansion. Indeed, for cosmological purposes,
we are curious for a source with $\rho\geq0$, in that case, all of
the energy conditions require $\omega\geq-1$ \cite{a}. The role of
possible DE candidates with $\omega<-1$ was pointed out by Caldwell,
who referred to null DEC violating sources as ''phantom''
components. It is argued that DE models with $w\geq-1$ such as the
cosmological constant and the quintessence satisfy the NEC, but the
models with $\omega<-1$ (predicted for instance by the phantom
theory), where the kinetic term of the scalar field has a wrong
(negative) sign, does not satisfy. However, quintom models can also
satisfy NEC as they yield the phantom era for a very short period of
time \cite{b}. Usually, the discussions on energy conditions for
cosmological constant are available in literature by introducing it
in some other type of matter like electromagnetic field \cite{c}.
The cosmological constant will trivially satisfy all these energy
conditions except SEC.

\section{Energy Conditions in the Most General Scalar-Tensor Gravity}

The most general scalar-tensor theory in 4-dimensions is given by
the action \cite{14,15}
\begin{eqnarray}\nonumber
S&=&\int
d^4x\sqrt{-g}[K(\phi,X)-G_{3}(\phi,X)\Box\phi+G_{4}(\phi,X)R+G_{4X}\{(\Box\phi)^2
\\\nonumber&-&(\nabla_\mu\nabla_{\nu}\phi)(\nabla^\mu\nabla^\nu\phi)\}
+G_{5}(\phi,X)G_{\mu\nu}(\nabla^\mu\nabla^\nu\phi)-\frac{1}{6}G_{5X}\{(\Box\phi)^3\\\nonumber
&-&3(\Box\phi)(\nabla_{\mu}\nabla_{\nu}\phi)(\nabla^\mu\nabla^\nu\phi)
+2(\nabla^\mu\nabla_{\alpha}\phi)(\nabla^\alpha\nabla_{\beta}\phi)(\nabla^\beta\nabla_{\mu}\phi)\}+L_{m}],\\\label{5}
\end{eqnarray}
where $\phi$ is the scalar field, $g$ is the determinant of the
metric tensor, $L_{m}$ denotes the matter part of the Lagrangian and
$X$ represents the kinetic energy term defined by
$X=-\frac{1}{2}\partial^\mu\phi\partial_{\mu}\phi$. Moreover,
$G_{\mu\nu}$ is the Einstein tensor, $R$ is the Ricci scalar,
$\nabla_{\mu}$ is the covariant derivative operator and
$\Box=\nabla_{\mu}\nabla^{\mu}$ is the de'Alembertian operator. The
functions $K(\phi,X)$ and $G_{i}(\phi,X);~i=3,4,5$ are all arbitrary
functions and $G_{iX}=\frac{\partial G_{i}}{\partial X}$. In this
action, the term $G_{3}(\phi,X)\Box\phi$ is the Galilean term,
$G_{4}(\phi,X)R$ can yield the Einstein-Hilbert term and
$G_{5}(\phi,X)$ leads to the interaction with Gauss-Bonnet term.
This indicates that it covers not only several DE proposals like
k-essence, $f(R)$ gravity, BD theory and Galilean gravity models but
it also contains 4-dimensional Dvali, Gabadadze and Poratti (DGP)
model (modified), the field coupling with Gauss-Bonnet term and the
field derivative coupling with Einstein tensor as its particular
cases.

By varying the action (\ref{5}) with respect to the metric tensor,
the gravitational field equation can be written as
\begin{eqnarray}\label{6}
G_{\mu\nu}=\frac{1}{G_{4}}\Theta_{\mu\nu}^{eff}=\frac{1}{G_{4}}[T^{m}_{\mu\nu}+T^{\phi}_{\mu\nu}],
\end{eqnarray}
where $\Theta_{\mu\nu}^{eff}$ is the modified energy-momentum
tensor, $T_{\mu\nu}^{m}$ is the source of usual matter field that
can be described by the perfect fluid, while $T_{\mu\nu}^{\phi}$
provides the matter source due to scalar field and hence yields the
source of DE, defined in Appendix \textbf{A}. The scalar wave
equation for such modified gravity has been described in literature
\cite{15}.

By inverting Eq.(\ref{6}), the Ricci tensor can be expressed in
terms of effective energy-momentum tensor and its trace as follows
\begin{eqnarray}\label{7}
R_{\mu\nu}=T_{\mu\nu}^{eff}-\frac{1}{2}g_{\mu\nu}T^{eff},
\end{eqnarray}
where the effective energy-momentum tensor $T^{eff}_{\mu\nu}$ and
its trace $T^{eff}$ are
\begin{eqnarray}\nonumber
T^{eff}_{\mu\nu}&=&T_{\mu\nu}+\frac{1}{2}K_{X}\nabla_\mu\phi\nabla_\nu\phi
-\frac{1}{2}G_{3X}\Box\phi\nabla_{\nu}\phi\nabla_{\mu}\phi
-\nabla_{(\mu}G_3\nabla_{\nu)}\phi+\frac{1}{2}G_{4X}\\\nonumber
&\times&R\nabla_{\mu}\phi\nabla_{\nu}\phi+\frac{1}{2}G_{4XX}[(\Box\phi)^2
-(\nabla_{\alpha}\nabla_{\beta}\phi)^2]\nabla_{\mu}\phi\nabla_{\nu}\phi+G_{4X}\Box\phi\\\nonumber
&\times&\nabla_{\mu}\nabla_{\nu}\phi
-G_{4X}\nabla_{\lambda}\nabla_\mu\phi\nabla^\lambda\nabla_\nu\phi-2\nabla_\lambda
G_{4X}\nabla^\lambda\nabla_{(\mu}\phi\nabla_{\nu)}\phi+\nabla_{\lambda}G_{4X}\\\nonumber
&\times&\nabla^\lambda\phi\nabla_\mu\nabla_\nu\phi-2[G_{4X}R_{\lambda(\mu}\nabla_{\nu)}
\phi\nabla^\lambda\phi-\nabla_{(\mu}G_{4X}\nabla_{\nu)}\phi\Box\phi]-G_{4X}\\\nonumber
&\times&R_{\mu\alpha\nu\beta}\nabla^\alpha\phi\nabla^\beta\phi
+G_{4\phi}\nabla_\mu\nabla_\nu\phi+G_{4\phi\phi}\nabla_\mu\phi\nabla_\nu\phi
-2G_{4X\phi}\nabla^\lambda\phi
\end{eqnarray}
\begin{eqnarray}\nonumber
&\times&\nabla_\lambda\nabla_{(\mu}\phi\nabla_{\nu)}\phi
+G_{4XX}\nabla^\alpha\phi\nabla_\alpha\nabla_\mu\phi\nabla^\beta\phi\nabla_\beta\nabla_\nu\phi
-G_{5X}R_{\alpha\beta}\nabla^\alpha\phi\\\nonumber
&\times&\nabla^\beta\nabla_{(\mu}\phi\nabla_{\nu)}\phi
+G_{5X}R_{\alpha(\mu}\nabla_{\nu)}\phi\nabla^\alpha\phi\Box\phi
+\frac{1}{2}G_{5X}R_{\alpha\beta}\nabla^\alpha\phi\nabla^\beta\phi\\\nonumber
&\times&\nabla_\mu\nabla_\nu\phi+\frac{1}{2}G_{5X}R_{\mu\alpha\nu\beta}\nabla^\alpha\phi\nabla^\beta\phi\Box\phi
-G_{5X}R_{\alpha\lambda\beta(\mu}\nabla_{\nu)}\phi\nabla^\lambda\phi
\nabla^\alpha\nabla^\beta\phi\\\nonumber
&-&G_{5X}R_{\alpha\lambda\beta(\mu}\nabla_{\nu)}\nabla^\lambda\phi
\nabla^\alpha\phi\nabla^\beta\phi+\frac{1}{2}\nabla_{(\mu}
[G_{5X}\nabla^\alpha\phi]\nabla_\alpha\nabla_{\nu)}\phi\Box\phi
-\frac{1}{2}\\\nonumber
&\times&\nabla_{(\mu}[G_{5\phi}\nabla_{\nu)}\phi]\Box\phi+\nabla_{\lambda}[G_{5\phi}\nabla_{(\mu}\phi]
\nabla_{\nu)}\nabla^\lambda\phi-\frac{1}{2}
[\nabla_\lambda(G_{5\phi}\nabla^\lambda\phi)\\\nonumber
&-&\nabla_\alpha(G_{5X}\nabla_\beta\phi)\nabla^\alpha\nabla^\beta\phi]\nabla_\mu\nabla_\nu\phi-\nabla^\alpha
G_{5}\nabla^\beta\phi
R_{\alpha(\mu\nu)\beta}+\nabla_{(\mu}G_{5}R_{\nu)\lambda}\\\nonumber
&\times&\nabla^\lambda\phi-\frac{1}{2}\nabla_{(\mu}G_{5X}\nabla_{\nu)}\phi[(\Box\phi)^2
-(\nabla_{\alpha}\nabla_{\beta}\phi)^2]+\nabla^\lambda
G_{5}R_{\lambda(\mu}\nabla_{\nu)}\phi \\\nonumber
&-&\nabla_\alpha[G_{5X}\nabla_\beta\phi]\nabla^\alpha\nabla_{(\mu}\phi\nabla^{\beta}\nabla_{\nu)}\phi+\nabla_\beta
G_{5X}[\Box\phi\nabla^\beta\nabla_{(\mu}\phi-\nabla^\alpha\nabla^\beta\phi\nabla_\alpha\\\nonumber
&\times&\nabla_{(\mu}\phi]\nabla_{\nu)}\phi
-\frac{1}{2}\nabla^\alpha\phi\nabla_{\alpha}G_{5X}[\Box\phi\nabla_{\mu}\nabla_\nu\phi-\nabla_{\beta}\nabla_\mu
\phi\nabla_{\beta}\nabla_\nu\phi]+\frac{1}{2}G_{5X}\\\nonumber&
\times&G_{\alpha\beta}\nabla^\alpha
\nabla^\beta\phi\nabla_\mu\phi\nabla_\nu\phi+\frac{1}{2}G_{5X}\Box\phi\nabla_\alpha
\nabla_\mu\phi\nabla^\alpha\nabla_\nu\phi-\frac{1}{2}G_{5X}(\Box\phi)^2\\\nonumber
&\times&\nabla_\mu\nabla_\nu\phi-\frac{1}{12}G_{5XX}[(\Box\phi)^3-3(\Box\phi)
(\nabla_\alpha\nabla_\beta\phi)^2+2(\nabla_\alpha\nabla_\beta\phi)^3]\nabla_\mu\phi\\\label{8}
&\times&\nabla_\nu\phi-\frac{1}{2}\nabla_\lambda
G_{5}R_{\mu\nu}\nabla^\lambda\phi,\\\nonumber
T^{eff}&=&K+\nabla_{\lambda}G_{3}\nabla^{\lambda}\phi-2(G_{4\phi}\Box\phi-2XG_{4\phi\phi})
-2\{-2G_{4X\phi}\nabla_\alpha\nabla_\beta\phi\nabla^\alpha\phi
\nabla^\beta\phi\\\nonumber
&+&G_{4XX}\nabla_\alpha\nabla_\lambda\phi\nabla_\beta\nabla^\lambda\phi\nabla^\alpha\phi\nabla^\beta\phi+
\frac{1}{2}G_{4X}[(\Box\phi)^2-(\nabla_\alpha\nabla_\beta\phi)^2]\}+2[G_{4X}\\\nonumber
&\times&R^{\alpha\beta}\nabla_\alpha\phi\nabla_\beta\phi-\nabla_\lambda
G_{4X}\nabla^\lambda\phi\Box\phi]-\frac{1}{2}R\nabla_\lambda
G_{5}\nabla^\lambda\phi-2\{-\frac{1}{6}G_{5X}[(\Box\phi)^3\\\nonumber
&-&3\Box\phi(\nabla_\alpha\nabla_\beta\phi)^2+2(\nabla_\alpha\nabla_\beta\phi)^3]+\nabla_\alpha
G_{5}R^{\alpha\beta}\nabla_\beta\phi-\frac{1}{2}\nabla_{\alpha}G_{5\phi}\nabla^\alpha\phi\Box\phi\\\nonumber
&+&\frac{1}{2}\nabla_\alpha
G_{5\phi}\nabla_\beta\phi\nabla^\alpha\nabla^\beta\phi-\frac{1}{2}\nabla_\alpha
G_{5X}\nabla^\alpha X\Box\phi+\frac{1}{2}\nabla_\alpha
G_{5X}\nabla_\beta X\nabla^\alpha\nabla^\beta\phi\\\nonumber
&-&\frac{1}{4}\nabla^\lambda
G_{5X}\nabla_\lambda\phi[(\Box\phi)^2-(\nabla_\alpha\nabla_\beta\phi)^2]
+\frac{1}{2}G_{5X}R_{\alpha\beta}\nabla^\alpha\phi\nabla^\beta\phi\Box\phi
-\frac{1}{2}G_{5X}\\\nonumber&\times&R_{\alpha\lambda\beta\rho}\nabla^\alpha\phi\nabla^\beta\phi\nabla^\lambda
\phi\nabla^\rho\phi\}+T+\frac{1}{2}K_{X}(\nabla_\mu\phi)^2
-\frac{1}{2}G_{3X}\Box\phi(\nabla_\mu\phi)^2\\\nonumber
&-&\nabla_\mu
G_{3}\nabla_\mu\phi+\frac{1}{2}G_{4X}R(\nabla_\mu\phi)^2+\frac{1}{2}G_{4XX}
[(\Box\phi)^2-(\nabla_\alpha\nabla_\beta\phi)^2](\nabla_\mu\phi)^2\\\nonumber
&+&G_{4X}\Box\phi\nabla_\mu\nabla_\mu\phi-G_{4X}\nabla_\lambda\nabla_\mu
\phi\nabla^\lambda\nabla_\mu\phi-2\nabla_\lambda
G_{4X}\nabla^\lambda(\nabla\phi)^2+\nabla_\lambda G_{4X}
\end{eqnarray}
\begin{eqnarray}\nonumber
&\times&\nabla^\lambda\phi\nabla_\mu\nabla_\mu\phi-2[G_{4X}R_{\lambda\mu}\nabla_\mu\phi\nabla^\lambda\phi
-\nabla_\mu G_{4X}\nabla_\mu\phi\Box\phi]\\\nonumber
&-&G_{4X}R_{\mu\alpha\mu\beta}\nabla^\alpha\phi\nabla^\beta\phi
+G_{4X}\nabla_\mu\nabla_\mu\phi+G_{4\phi\phi}(\nabla_\mu\phi)^2
-2G_{4X\phi}\nabla^\lambda\phi\\\nonumber
&\times&\nabla_\lambda(\nabla_{\mu}\phi)^2+G_{4XX}\nabla^\alpha\phi\nabla_\alpha
\nabla_\mu\phi\nabla^\beta\phi\nabla_\beta\nabla_\mu\phi
-G_{5X}R_{\alpha\beta}\nabla^\alpha\phi\nabla^\beta(\nabla_{\mu}\phi)^2\\\nonumber
&+&G_{5X}R_{\alpha(\mu}\nabla_{\mu)}\phi\nabla^\alpha\phi\Box\phi
+\frac{1}{2}G_{5X}R_{\alpha\beta}\nabla^\alpha\phi\nabla^\beta\phi\nabla_\mu\nabla_\mu\phi
+\frac{1}{2}G_{5X}\\\nonumber
&\times&R_{\mu\alpha\mu\beta}\nabla^\alpha\phi\nabla^\beta\phi\Box\phi
-G_{5X}R_{\alpha\lambda\beta\mu}\nabla_{\mu}\phi\nabla^\lambda\phi
\nabla^\alpha\nabla^\beta\phi-G_{5X}R_{\alpha\lambda\beta\mu}\\\nonumber
&\times&\nabla_{\mu}\nabla^\lambda\phi
\nabla^\alpha\phi\nabla^\beta\phi+\frac{1}{2}\nabla_{(\mu}[G_{5X}\nabla^\alpha\phi]
\nabla_\alpha\nabla_{\mu)}\phi\Box\phi
-\frac{1}{2}\nabla_{(\mu}[G_{5\phi}\nabla_{\mu)}\phi]\Box\phi\\\nonumber
&+&\nabla_{\lambda}[G_{5\phi}\nabla_{(\mu}\phi]\nabla_{\mu)}\nabla^\lambda\phi-\frac{1}{2}
[\nabla_\lambda(G_{5\phi}\nabla^\lambda\phi)-\nabla_\alpha(G_{5X}\nabla_\beta\phi)
\nabla^\alpha\nabla^\beta\phi]\\\nonumber
&\times&\nabla_\mu\nabla_\mu\phi-\nabla^\alpha G_{5}\nabla^\beta\phi
R_{\alpha\mu\mu\beta}+\nabla_{\mu}G_{5}R_{\mu\lambda}\nabla^\lambda\phi
-\frac{1}{2}\nabla_{\mu}G_{5X}\\\nonumber
&\times&\nabla_{\mu}\phi[(\Box\phi)^2
-(\nabla_{\alpha}\nabla_{\beta}\phi)^2]+\nabla^\lambda
G_{5}R_{\lambda\mu}\nabla_{\mu}\phi-\nabla_\alpha[G_{5X}\nabla_\beta\phi]
\nabla^\alpha\nabla_{\mu}\phi\\\nonumber
&\times&\nabla^{\beta}\nabla_{\mu}\phi+\nabla_\beta
G_{5X}[\Box\phi\nabla^\beta\nabla_{(\mu}\phi-\nabla^\alpha\nabla^\beta
\phi\nabla_\alpha\nabla_{(\mu}\phi]\nabla_{\mu)}\phi
-\frac{1}{2}\nabla^\alpha\phi\\\nonumber
&\times&\nabla_{\alpha}G_{5X}[\Box\phi\nabla_{\mu}\nabla_\mu\phi-\nabla_{\beta}\nabla_\mu
\phi\nabla_{\beta}\nabla_\mu\phi]+\frac{1}{2}G_{5X}G_{\alpha\beta}\nabla^\alpha
\nabla^\beta\phi(\nabla_\mu\phi)^2\\\nonumber
&+&\frac{1}{2}G_{5X}\Box\phi\nabla_\alpha\nabla_\mu\phi\nabla^\alpha\nabla_\mu\phi-\frac{1}{2}
G_{5X}(\Box\phi)^2\nabla_\mu\nabla_\mu\phi-\frac{1}{12}G_{5XX}[(\Box\phi)^3\\\nonumber
&-&3(\Box\phi)(\nabla_\alpha\nabla_\beta\phi)^2+2(\nabla_\alpha\nabla_\beta\phi)^3](\nabla_\mu\phi)^2
-\frac{1}{2}\nabla_\lambda G_{5}R\nabla^\lambda\phi \\\nonumber
&+&\frac{1}{2}\nabla_{\alpha}G_{5X}\nabla_{\beta}X\nabla^\alpha\nabla^\beta\phi-\frac{1}{4}\nabla^\alpha
G_{5X}\nabla_\lambda\phi[(\Box\phi)^2-(\nabla_\alpha\nabla_\beta\phi)^2]\\\label{9}
&+&\frac{1}{2}G_{5X}R_{\alpha\beta}\nabla^\alpha\phi\nabla^\beta\phi\Box\phi
-\frac{1}{2}G_{5X}R_{\alpha\lambda\beta\rho}.
\end{eqnarray}
Evaluating the temporal and spatial components of the effective
energy-momentum tensor and its trace defined above and using these
values in Eqs.(\ref{1*}) and (\ref{2*}), we can find the energy
conditions for any spacetime.

Let us consider the spatially homogeneous, isotropic and flat FRW
universe model with $a(t)$ as a scale factor described by the metric
\begin{eqnarray}\label{10}
ds^2=-dt^2+a^2(t)(dx^2+dy^2+dz^2).
\end{eqnarray}
The background fluid is taken as perfect fluid given by
Eq.(\ref{3}) with $u_\mu=(1,0,0,0)$ and the null like vector is
taken as $k_\mu=(1,a,0,0)$. Furthermore, we assume that the scalar
field is a function of time only. The Friedmann equations for the
generalized scalar-tensor theory in terms of effective energy
density and pressure are given by \cite{15}
\begin{eqnarray}\label{11}
3H^2=\frac{\rho^{eff}}{G_{4}}, \quad
-(3H^2+2\dot{H})=\frac{p^{eff}}{G_{4}},
\end{eqnarray}
where
\begin{eqnarray}\nonumber
\rho^{eff}&=&\frac{1}{2}[\rho^{m}+2XK_{X}-K+6X\dot{\phi}HG_{3X}
-2XG_{3\phi}+24H^2X(G_{4X}\\\nonumber
&+&XG_{4XX})-12HX\dot{\phi}G_{4X\phi}-6H\dot{\phi}G_{4\phi}
+2H^3X\dot{\phi}(5G_{5X}+2XG_{5XX})\\\label{12}
&-&6H^2X(3G_{5\phi}+2XG_{5X\phi})],\\\nonumber
p^{eff}&=&\frac{1}{2}[p^{m}+K-2X(G_{3\phi}+\ddot{\phi}G_{3X})
-12H^2XG_{4X}-4H\dot{X}G_{4X}\\\nonumber
&-&8\dot{H}XG_{4X}-8HX\dot{X}G_{4XX}+2(\ddot{\phi}+2H\dot{\phi})
G_{4\phi}+4XG_{4\phi\phi}+4X(\ddot{\phi}
\end{eqnarray}
\begin{eqnarray}\nonumber
&-&2H\dot{\phi})G_{4X\phi}-2X(2H^3\dot{\phi}+2H\dot{H}\dot{\phi}
+3H^2\ddot{\phi})G_{5X}-4H^2X^2\ddot{\phi}G_{5XX}\\\nonumber
&+&4HX(\dot{X}-HX)G_{5X\phi}+2[2(HX\dot{)}+3H^2X]G_{5\phi}
+4HX\dot{\phi}G_{5\phi\phi}].\\\label{13}
\end{eqnarray}
Here $G_4$, being an arbitrary function of $\phi$ and $X$, acts as a
dynamical gravitational constant and it should be positive for any
gravitational theory. Furthermore, $\rho^m$ and $p^m$ are density
and pressure, respectively for ordinary matter. We shall discuss its
different forms in the next section. Using these values in
Eq.(\ref{4}), it can be checked that the NEC, WEC, SEC and DEC
require the following conditions to be satisfied,
\begin{eqnarray}\nonumber
&&\textbf{NEC}:\quad\rho^{eff}+p^{eff}\geq0\quad\Rightarrow\\\nonumber
&&\frac{1}{2G_{4}}[(\rho^{m}+p^{m})+2XK_{X}+6X\dot{\phi}HG_{3X}
-4XG_{3\phi}+12H^2XG_{4X}\\\nonumber
&&+24H^2X^2G_{4XX}-20XH\dot{\phi}G_{4X\phi}-2H\dot{\phi}G_{4\phi}
+6H^3X\dot{\phi}G_{5X}+4H^3X^2\dot{\phi}\\\nonumber &&\times
G_{5XX}-16H^2X^2G_{5X\phi}-12H^2XG_{5\phi}-2X\ddot{\phi}G_{3X}
-4H\dot{X}G_{4X}-8\dot{H}X\\\nonumber&&\times
G_{4X}-8H\dot{X}XG_{4XX}+2\ddot{\phi}G_{4\phi}+4XG_{\phi\phi}+4X\ddot{\phi}
G_{4X\phi}-2X(2H\dot{H}\dot{\phi}\\\nonumber
&&+3H^2\ddot{\phi})G_{5X}-4H^2X^2\ddot{\phi}
G_{5XX}+4HX\dot{X}G_{5X\phi}+4(HX\dot{)}G_{5\phi}\\\label{14}
&&+4HX\dot{\phi}G_{5\phi\phi}]\geq0,\\\nonumber
&&\textbf{WEC}:\quad\rho^{eff}+p^{eff}\geq0,\quad\rho^{eff}\geq0\quad\Rightarrow\\\nonumber
&&\frac{1}{2G_{4}}[\rho^{m}+2XK_{X}
-K+6XH\dot{\phi}G_{3X}-2XG_{3\phi}+24H^2X\\\nonumber
&&\times(G_{4X}+XG_{4XX})-12XH\dot{\phi}-6H\dot{\phi}G_{4\phi}+
2H^3X\dot{\phi}(5G_{5X}+2XG_{5XX})\\\label{15}
&&-6H^2X(3G_{5\phi}+2XG_{5X\phi})]\geq0,
\end{eqnarray}
\begin{eqnarray}\nonumber
&&\textbf{SEC}:\quad\rho^{eff}+p^{eff}\geq0,\quad\rho^{eff}+3p^{eff}\geq0\quad\Rightarrow\\\nonumber
&&\frac{1}{2G_{4}}[(\rho^{m}+3p^{m})+2XK_{X}+2K+6XH\dot{\phi}G_{3X}
-8XG_{3\phi}-6X\ddot{\phi}G_{3X}\\\nonumber
&&-12H^2XG_{4X}+24H^2X^2G_{4XX}-36XH\dot{\phi}G_{4X\phi}
+6H\dot{\phi}G_{4\phi}+6\ddot{\phi}G_{4\phi}\\\nonumber
&&-2H^3X\dot{\phi}G_{5X}+4H^3X^2\dot{\phi}G_{5XX}-24H^2X^2G_{5X\phi}
+12HX\dot{X}G_{5X\phi}\\\nonumber
&&+12(HX\dot{)}G_{5\phi}+12HX\dot{\phi}G_{\phi\phi}-12H^2X^2\ddot{\phi}
G_{5XX}-12XH\dot{H}\dot{\phi}G_{5X}\\\nonumber
&&-18XH^2\ddot{\phi}G_{5X}
+12X\ddot{\phi}G_{4X\phi}-24HX\dot{\phi}G_{4X\phi}+12XG_{4\phi\phi}-24HX\dot{X}\\\label{16}
&&\times
G_{4XX}-24\dot{H}XG_{4X}-12H\dot{X}G_{4X}-36H^2XG_{4X}]\geq0,\\\nonumber
&&\textbf{DEC}:\quad\rho^{eff}\geq0,\quad\rho^{eff}+p^{eff}\geq0,
\quad\rho^{eff}-p^{eff}\geq0\quad\Rightarrow\\\nonumber
&&\frac{1}{2G_{4}}[(\rho^{m}-p^{m})+2XK_{X}-2K+6XH\dot{\phi}
G_{3X}+2X\ddot{\phi}G_{3X}+36H^2XG_{4X}\\\nonumber
&&+24H^2XG_{4XX}-4XH\dot{\phi}
G_{4X\phi}-10H\dot{\phi}G_{4\phi}+14H^3X\dot{\phi}G_{5X}
+4H^3X^2\dot{\phi}\\\nonumber &&\times
G_{5XX}-24H^2XG_{5\phi}-4(HX\dot{)}G_{5\phi}-4HX\dot{\phi}G_{\phi\phi}
-8H^2X^2G_{5X\phi}+4H\dot{X}\\\nonumber &&\times
G_{4X}+8\dot{H}XG_{4X}+8HX\dot{X}G_{4XX}-2\ddot{\phi}G_{4\phi}
-4XG_{4\phi\phi}-4X\ddot{\phi}G_{4X\phi}\\\label{17}
&&+2X(2H\dot{H}\dot{\phi}+3H^2\ddot{\phi})G_{5X}+4H^2X^2G_{5XX}\ddot{\phi}
-4HX\dot{X}G_{5X\phi}]\geq0.
\end{eqnarray}

In a mechanical framework, the terms velocity, acceleration, jerk
and snap parameters are based on the first four time derivatives of
position. In cosmology, the Hubble, deceleration, jerk and snap
parameters are, respectively, defined as
\begin{eqnarray}\label{18}
H=\frac{\dot{a}}{a},\quad q=-\frac{1}{H^2}\frac{\ddot{a}}{a},\quad
j=\frac{1}{H^3}\frac{\dddot{a}}{a}, \quad
s=\frac{1}{H^4}\frac{\ddddot{a}}{a}.
\end{eqnarray}
In order to have a more precise picture of these conditions
(\ref{14})-(\ref{17}), we use the relations of time derivatives of
Hubble parameter in terms of cosmological quantities like
deceleration, snap and jerk parameters as
\begin{eqnarray}\label{19}
\dot{H}=-H^2(1+q), \quad \ddot{H}=H^3(j+3q+2), \quad
\dddot{H}=H^4(s-2j-5q-3).
\end{eqnarray}
Moreover, we assume that the scalar field evolves as a power of
scale factor, i.e., $\phi(t)\sim a(t)^\beta$ \cite{**}, which leads
to
\begin{eqnarray}\label{20}
&&\dot{\phi}\sim \beta Ha^\beta,\quad \ddot{\phi}\sim\beta
a^\beta(\dot{H}+\beta H^2)=\beta a^\beta H^2(\beta-1-q),\\\label{21}
&&X\sim\frac{\beta^2H^2a^{2\beta}}{2},\quad\dot{X}\sim\beta^2H^3a^{2\beta}(\beta-1-q).
\end{eqnarray}
Here $\beta$ ia a non-zero parameter ($\beta=0$ yields constant
scalar field, $\beta>0$ yields expanding scalar field and $\beta<0$
corresponds to contracting scalar field). Clearly, $\phi$ remains as
a positive quantity.

Introducing these quantities in the energy conditions given by
Eqs.(\ref{14})-(\ref{17}), it follows that
\begin{eqnarray}\nonumber
&&\frac{1}{G_4}[K_{X}-2G_{3,\phi}+6H^2G_{4X}-6H^2G_{5\phi}+2G_{4\phi\phi}]\beta^2H^2a^{2\beta}+[3HG_{3X}\\\nonumber
&&-10HG_{4X\phi}+3H^3G_{5X}+2HG_{5\phi\phi}]\beta^3H^3a^{3\beta}+(6H^2G_{4XX}-4H^2G_{5X\phi})\\\nonumber
&&\times\beta^4H^4a^{4\beta}-2H^2\beta a^\beta
G_{4\phi}+H^8\beta^5a^{5\beta}G_{5XX}-\beta^3H^4a^{3\beta}(\beta-1-q)G_{3X}\\\nonumber
&&-4H^4a^{2\beta}\beta^2(\beta-1-q)G_{4X}
+4H^4a^2\beta\beta^2(1+q)G_{4X}-4H^6\beta^4a^{4\beta}(\beta\\\nonumber
&&-1-q)G_{4XX}+2H^2a^\beta(\beta-1-q)\beta
G_{4\phi}+2\beta^3H^4a^{3\beta}(\beta-1-q)G_{4X\phi}\\\nonumber
&&+2H^6(1+q)\beta^3a^{3\beta}G_{5X}-3H^6\beta^3a^{3\beta}(\beta-1-q)G_{5X}-H^8\beta^5a^{5\beta}
(\beta-1\\\nonumber
&&-q)G_{5XX}+2H^6\beta^4a^{4\beta}(\beta-1-q)G_{5X\phi}+4H^4\beta^2a^{2\beta}
(\beta-1-q)G_{5\phi}\\\label{22}
&&-2H^4(1+q)\beta^2a^{2\beta}G_{5\phi}+(\rho^{m}+p^{m})\geq0,\\\nonumber
&&\frac{1}{G_4}[\rho^{m}+3p^{m}+2K+3\beta^3a^{3\beta}H^4G{3X}-4\beta^2H^2a^{2\beta}
G_{3\phi}-3\beta^3a^{3\beta}H^4(\beta-1\\\nonumber
&&-q)G_{3X}-6H^4\beta^2a^{2\beta}G_{4X}+3H^6\beta^4a^{4\beta}G_{4XX}
-18H^4\beta^3a^{3\beta}G_{X\phi}+6H^2\beta\\\nonumber &&\times
a^\beta G_{4\phi}+6H^2\beta
a^\beta(\beta-1-q)G_{4\phi}-H^6\beta^3a^{3\beta}G_{5X}+2H^8\beta^{5}
a^{5\beta}G_{5XX}\\\nonumber
&&-3H^6\beta^4a^{4\beta}G_{5X\phi}+6H^6\beta^4a^{4\beta}(\beta-1-q)
G_{5X\phi}-6H^4\beta^2a^{2\beta}(1+q)G_{5\phi}\\\nonumber
&&+12H^4\beta^2a^{2\beta}(\beta-1-q)G_{5\phi}+6H^4\beta^3a^{3\beta}
G_{5\phi\phi}-3H^8\beta^5a^{5\beta}(\beta-1-q)\\\nonumber &&\times
G_{5XX}+H^6\beta^3a^{3\beta}(1+q)G_{5X}-9H^6\beta^3a^{3\beta}(\beta-1-q)
G_{5X}+6H^4\beta^3a^{3\beta}\\\nonumber&&\times
(\beta-1-q)G_{4X\phi}-12H^4\beta^3a^{3\beta}G_{4X\phi}+6H^2\beta^2a^{2\beta}G_{4\phi\phi}
-12H^6\beta^4a^{4\beta}\\\nonumber&&\times(\beta-1-q)G_{4XX}+12H^4\beta^2a^{2\beta}(1+q)G_{4X}
-12H^4\beta^2a^{2\beta}(\beta-1-q)G_{4X}\\\label{23}
&&-18H^4\beta^2a^{2\beta}G_{4X}+\beta^2a^{2\beta}H^2K_{X}]\geq0,\\\nonumber
&&\frac{1}{G_4}[H^2\beta^2a^{2\beta}K_{X}-K+3H^4\beta^3a^{3\beta}G_{3X}-H^2\beta^2a^{2\beta}G_{3\phi}
+12H^4\beta^2a^{2\beta}G_{4X}\\\nonumber
&&+6H^6\beta^4a^{4\beta}G_{4XX}-6H^4\beta^3a^{3\beta}G_{X\phi}-6H^2\beta
a^\beta G_{4\phi}+5H^6\beta^3a^{3\beta}G_{5X}\\\label{24}
&&+H^8\beta^5a^{5\beta}G_{5XX}-9H^4\beta^2a^{2\beta}G_{5\phi}
-3H^6\beta^4a^{4\beta}G_{4X\phi}+\rho^m]\geq0,\\\nonumber
&&\frac{1}{G_4}[\rho^m-p^m+\beta^2H^2a^{2\beta}K_{X}-2K+3H^4\beta^3a^{3\beta}G_{3X}
+\beta^2H^3a^{2\beta}(\beta-q-1)\\\nonumber &&\times
G_{3X}+18\beta^2H^4a^{2\beta}G_{4X}+12H^4\beta^2a^{2\beta}G_{4XX}
-2H^4\beta^3a^{3\beta}G_{4X\phi}-10H^2\\\nonumber &&\times\beta
a^\beta
G_{4\phi}+7H^6\beta^3a^{3\beta}G_{5X}+H^8\beta^5{a^5\beta}G_{5XX}-12H^4\beta^2a^{2\beta}G_{5\phi}
-4H^4\beta^2\\\nonumber&&\times
a^{2\beta}G_{5\phi}+2H^4\beta^2a^{2\beta}(1+q)G_{5\phi}-2H^4\beta^3a^{3\beta}G_{5\phi\phi}
-4H^4\beta^2a^{2\beta}G_{X\phi}+4\\\nonumber &&\times
H^4\beta^2\times a^{2\beta}(\beta-q-1)G_{4X}-
4H^4\beta^2a^{2\beta}(1+q)G_{4X}+4H^6\beta^4a^{4\beta}(\beta-1
\end{eqnarray}
\begin{eqnarray}\nonumber
&&-q)G_{4XX}-2H^2a^\beta\beta(\beta-1-q)G_{4\phi}-2H^2\beta^2a^{2\beta}G_{4\phi\phi}
-2H^4\beta^3a^{3\beta}(\beta-q\\\nonumber
&&-1)G_{4X\phi}-2H^6\beta^3a^{3\beta}(1+q)G_{5X}+3H^6\beta^3a^{3\beta}(\beta-1-q)G_{5X}+H^8\beta^5\\\label{25}
&&\times
a^{5\beta}(\beta-1-q)G_{5XX}-2H^6\beta^4a^{4\beta}(\beta-1-q)G_{5X\phi}]\geq0.
\end{eqnarray}
These are the most general energy conditions that can yield the
energy conditions for various DE models like k-essence and modified
theories in certain limits. In order to satisfy these conditions, it
must be guaranteed that the function $G_{4}$ is a positive quantity.
However, we have discussed earlier that $G_4$, being a gravitational
constant, would be positive in all cases (if it is not so, then we
impose this condition and restrict the free parameters). Clearly,
these conditions are only dependent on the Hubble and deceleration
parameters as well as arbitrary functions namely $K,~G_{3},~G_{4}$
and $G_{5}$. Once these arbitrary functions are specified, the
energy bounds on the selected models can be determined by using
these conditions.

In order to have a better understanding of these constraints, we can
use either the power law anstaz for the scale factor, e.g.,
\cite{22} or we can use the estimation of present values of the
respective parameters available in literature. In this study, we
consider the present value of the Hubble parameter $H_{0}=0.718$,
the scale factor $a_{0}=1$ and the deceleration parameter $q=-0.64$
as suggested by Cappozzielo et al. \cite{24}. Since it is well-known
that the energy constraints are satisfied for usual matter contents
like perfect fluid, therefore we shall focus on validity of the
energy constraints for the scalar field terms only (either we take
vacuum case or assume that the energy conditions for ordinary matter
hold). It is interesting to mention here that the respective energy
conditions in GR can be recovered by taking the arbitrary functions
$K,~G_{3},$ and $G_{5}$ zero with $G_{4}$ as constant.

\section{Energy Conditions in Some Particular Cases}

Now we discuss application of the derived conditions to some
particular cases of this theory. The violation of energy conditions
leads to various interesting results. In particular, for a canonical
scalar field, violation of these conditions yields instabilities and
ghost pathologies. It is important to discuss the violation of these
energy conditions in order to check the existence of instabilities
in Horndeski theory. The procedure for FRW universe model in most
general scalar-tensor theory based on tensor and scalar
perturbations is available in literature \cite{14}. By introducing
perturbed metric, it has been shown that for the avoidance of ghost
and gradient instabilities, the tensor perturbations suggest
\begin{eqnarray}\label{26*}
\mathcal{F}_T&=&2[G_4-X(\ddot{\phi}G_{5X}+G_{5\phi})]>0,\\\label{26**}
\mathcal{G}_{T}&=&2[G_4-2XG_{4X}-X(H\dot{\phi}G_{5X}-G_{5\phi})]>0,
\end{eqnarray}
while scalar perturbations impose
\begin{eqnarray}\label{27*}
\mathcal{F}_S&=&\frac{1}{a}\frac{d}{dt}(\frac{a}{\Theta}\mathcal{G}_T^2)-\mathcal{F}_T>0,\\\label{27**}
\mathcal{G}_S&=&\frac{\Sigma}{\Theta^2}\mathcal{G}_T^2+3\mathcal{G}_T>0,
\end{eqnarray}
where the quantities $\Sigma$ and $\Theta$ are defined in \cite{14}.
We simply plug the values in these conditions for the following
cases and show that violation of energy conditions leads to the
existence of ghost instabilities.

\subsection{k-essence Models in General Relativity}

The k-essence dynamical models of DE play a dominant role in the
solution of various problems in cosmological context \cite{25}.
The action (\ref{5}) can be reduced to the action for k-essence
model in GR framework defined by $S=\int
\sqrt{-g}[K(\phi,X)+\frac{M_{pl}^2}{2}R+L_{m}]d^4x$ with the
following choice of the functions
\begin{eqnarray}\label{26}
K=K(\phi,X),\quad G_{4}=\frac{M_{pl}^2}{2},\quad G_{3}=G_{5}=0,
\end{eqnarray}
where $M_{pl}^2$ is Planck mass. The k-essence models can be
classified into three forms:
\begin{itemize}
\item $K(\phi,X)=K_{1}(X)$ (Kinetic case),
\item $K(\phi,X)=K_1(X)B(\phi)$,
\item $K(\phi,X)=K_1(X)+B(\phi)$.
\end{itemize}
For the choice of arbitrary functions given by Eq.(\ref{26}), the
energy conditions (\ref{14})-(\ref{17}) take the following forms
\begin{eqnarray}\nonumber
\textbf{NEC}:\quad\frac{1}{M_{pl}^2}[2XK_X+\rho^m+p^m]\geq0,
\end{eqnarray}
\begin{eqnarray}\nonumber &&\textbf{WEC}:\quad\rho^{eff}+p^{eff}\geq0,
\quad\frac{1}{M_{pl}^2}[2XK_X-K+\rho^m]\geq0,\\\nonumber
&&\textbf{SEC}:\quad\rho^{eff}+p^{eff}\geq0,
\quad\frac{1}{M_{pl}^2}[2XK_X+2K+\rho^m+3p^m]\geq0,\\\nonumber
&&\textbf{DEC}:\quad\rho^{eff}+p^{eff}\geq0,
\quad\rho^{eff}\geq0,\quad\frac{1}{M_{pl}^2}[2XK_X-2K+\rho^m-p^m]\geq0.
\end{eqnarray}
Here $K(\phi,X)$ is arbitrary.

In order to see how the function $K(\phi,X)$ can be constrained by
using the above energy conditions, we choose a particular model of
k-essence \cite{26}
\begin{eqnarray}\nonumber
K(\phi,X)=\frac{1}{2}[C_2-C_1-2B\phi^2]+\frac{1}{2}(C_1+C_2)X+\frac{1}{2}M_{0}^4(X-1)^2,
\end{eqnarray}
where $C_1,~C_2,~B$ and $M_{0}$ are arbitrary constants. In this
case, WEC requires the following conditions
\begin{eqnarray}\nonumber
&&\frac{1}{M_{pl}^2}[\frac{1}{2}X(C_1+C_2)+2XM_{0}^4(X-1)-\frac{1}{2}M_{0}^4(X-1)^2-\frac{1}{2}(C_2\\\nonumber
&&-C_1-2B\phi^2)+\rho^m]\geq0,\\\label{27}
&&\frac{1}{M_{pl}^2}[X(C_1+C_2)+2M_{0}^4X(X-1)+\rho^m+p^m]\geq0,
\end{eqnarray}
where $M_{pl}^2>0$. For the interpretation of the above
inequalities, we consider the power law ansatz for the scalar
field $\phi\sim a^\beta;~\beta\neq0$, which further yields
$X\sim\frac{\beta^2a^{2\beta}H^2}{2}$. Consequently, the WEC
(\ref{27}) turns out to be
\begin{eqnarray}\nonumber
&&[\frac{1}{2}\frac{\beta^2a^{2\beta}H^2}{2}(C_1+C_2)+2\frac{\beta^2a^{2\beta}H^2}{2}
M_{0}^4(\frac{\beta^2a^{2\beta}H^2}{2}-1)
-\frac{1}{2}M_{0}^4(\frac{\beta^2a^{2\beta}H^2}{2}\\\label{28}
&&-1)^2-\frac{1}{2}(C_2-C_1-2B\phi^2)+\rho^m]\geq0,\\\label{29}
&&[\frac{\beta^2a^{2\beta}H^2}{2}(C_1+C_2)+2M_{0}^4\frac{\beta^2a^{2\beta}H^2}{2}
(\frac{\beta^2a^{2\beta}H^2}{2}-1)+\rho^m+p^m]\geq0.
\end{eqnarray}
\begin{figure}
\centering \epsfig{file=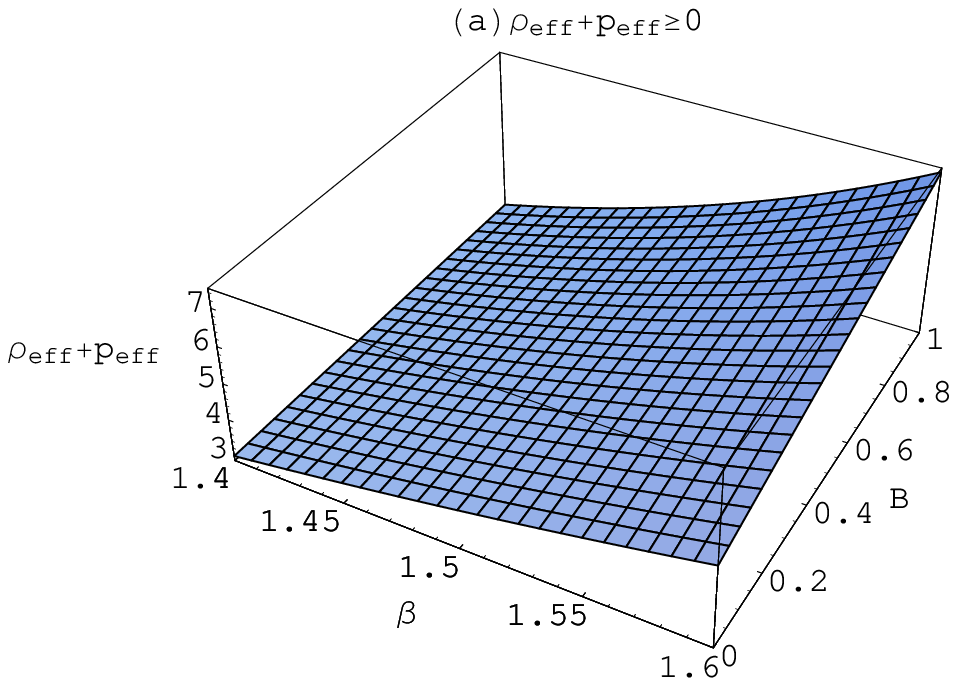,width=.45\linewidth}
\epsfig{file=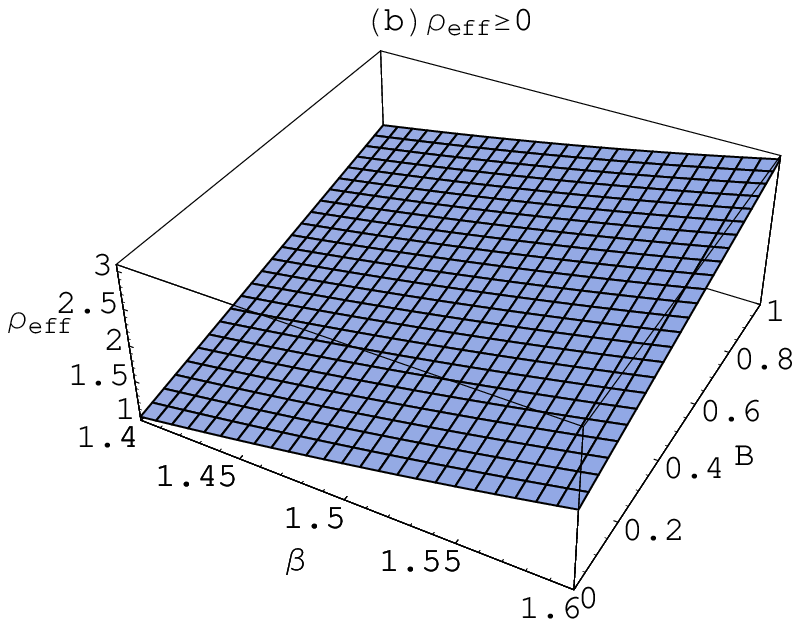,width=.45\linewidth} \caption{Plots (a)
shows $\rho^{eff}+p^{eff}\geq0$ versus $\beta$ and $M_0$. Plot (b)
represents $\rho^{eff}\geq0$ versus parameters $\beta$ and $B$
with $M_{0}=0.2$. In both cases, we take the present values of
cosmological parameters with $C_1=1,~C_2=2$.}
\end{figure}

It is difficult to find the admissible ranges of all constants
$C_1,~C_2,~B,~M_{0}$ and $\beta$ from the above conditions. In order
to find the constraints on these parameters, we consider that these
conditions are satisfied for ordinary matter, i.e., $\rho^m>0$ and
$\rho^m+p^m>0$. Moreover, we take the present value of Hubble
parameter and choose some particular values of the constants $C_1$
and $C_2$ to find the ranges of $B,~M_0$ and $\beta$, consistent
with the WEC. It turns out from the graphs that we can take the
parameter $\beta$ as follows $\beta>1.4,~0<\beta<1.4$ and $\beta<0$,
while $B$ and can be positive or negative. For the consistency of
condition (\ref{29}), we restrict the parameter $M_0$ as $0<M_0<1$.
From the condition (\ref{28}), it can be observed that energy
conditions are satisfied only when we take $\beta>1.4$ with
arbitrary $B$ and $\beta<1.4$ with $B>10$ only. Other choices of
these parameters lead to violation of WEC. Figures \textbf{1(a)} and
\textbf{(b)} show that WEC is satisfied with these fixed input
parameters by taking $\beta>1.4,~0<M_0<1$ and $0<B<1$.

\subsection{Brans-Dicke Theory}

Brans-Dicke gravity with action
$\int\sqrt{-g}[\frac{M_{pl}X\omega}{\phi}-V(\phi)+\frac{1}{2}M_{pl}R\phi+L_{m}]d^4x$
can be defined by the following choice of functions
\begin{eqnarray}\label{30}
K=\frac{M_{pl}X\omega}{\phi}-V(\phi),\quad G_{3}=G_{5}=0,\quad
G_{4}=\frac{1}{2}M_{pl}\phi.
\end{eqnarray}
Here $\omega$ is the BD parameter and $V$ is the field potential.
The action for general scalar-tensor gravity can be obtained by
taking $F(\phi)$ instead of $\phi$ in $G_4$. In this case, the
energy conditions take the form
\begin{eqnarray}\nonumber
\textbf{NEC}:\quad&&\frac{1}{M_{pl}}[\frac{\dot{\phi}^2}{\phi^2}M_{pl}\omega-\frac{H\dot{\phi}}{\phi}M_{pl}
+\frac{\ddot{\phi}}{\phi}M_{pl}+\frac{\rho^m+p^m}{\phi}]\geq0,\\\nonumber
\textbf{WEC}:\quad&&\rho^{eff}+p^{eff}\geq0,
\quad\frac{1}{M_{pl}}[\frac{\omega
M_{pl}\dot{\phi}^2}{2\phi^2}+\frac{V(\phi)}{\phi}-\frac{3H\dot{\phi}}{\phi}M_{pl}+\frac{\rho^m}{\phi}]\geq0,
\\\nonumber
\textbf{SEC}:\quad&&\rho^{eff}+p^{eff}\geq0,
\quad\frac{1}{M_{pl}}[2\frac{\dot{\phi}^2}{\phi^2}M_{pl}
\omega-\frac{2V(\phi)}{\phi}+\frac{3H\dot{\phi}}{\phi}M_{pl}
+3\frac{\ddot{\phi}}{\phi}M_{pl}\\\nonumber
&&+\frac{\rho^m+3p^m}{\phi}]\geq0,\\\nonumber
\textbf{DEC}:\quad&&\rho^{eff}+p^{eff}\geq0,\quad\rho^{eff}\geq0,
\quad\frac{1}{M_{pl}}[2\frac{V(\phi)}{\phi}-5\frac{H\dot{\phi}}
{\phi}M_{pl}\\\nonumber
&&-\frac{\ddot{\phi}}{\phi}M_{pl}+\frac{\rho^m-p^m}{\phi}]\geq0.
\end{eqnarray}
A suitable choice of the field potential has always been a matter
of debate in BD gravity. We use power laws for the scalar field
and potential as $\phi=\phi_{0}a^\beta$ and $V=V_{0}\phi^m$,
respectively where $m$ and $\beta$ are non zero parameters.
Consequently, the WEC restricts the parameters as
\begin{eqnarray}\label{31}
V_{0}\geq3H_{0}^2\beta-\frac{\omega}{2}\beta^2H_{0}^2,\quad
\beta^2H_{0}^2\omega-H_{0}^2\beta+H_{0}^2a^\beta\beta^2-a^\beta\beta
H_{0}^2(1+q_0)\geq0.
\end{eqnarray}
\begin{figure}
\centering \epsfig{file=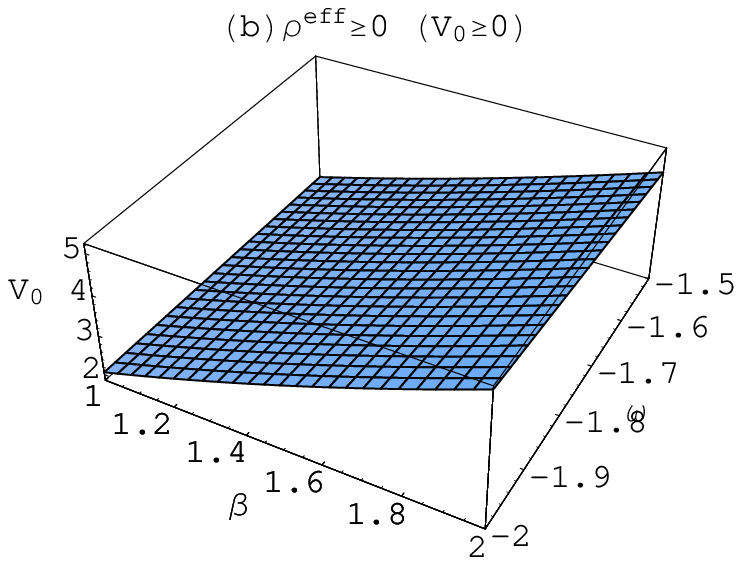,width=.45\linewidth}
\epsfig{file=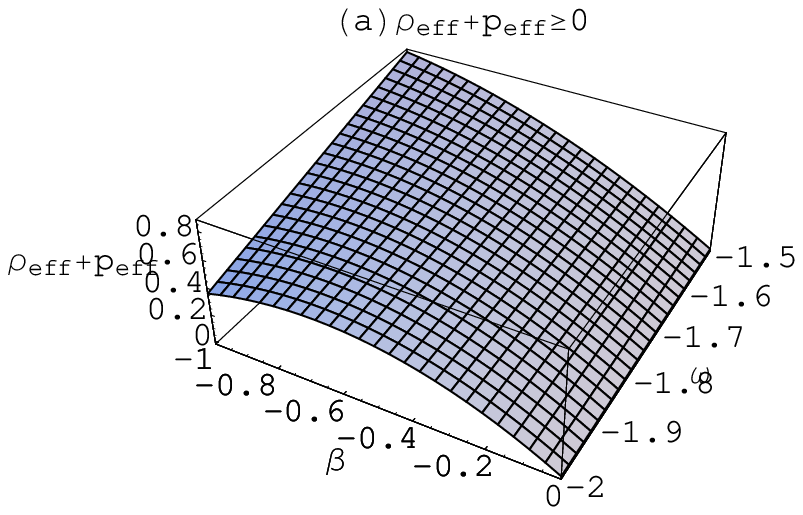,width=.45\linewidth} \caption{Plots (a) and
(b) show WEC versus parameters $\omega$ and $\beta$ for $H_0=0.718$
and $q_0=-0.64$.}
\end{figure}

For accelerated expanding universe, the observed range of BD
parameter is $-2<\omega<-3/2$ \cite{26*}. Clearly, both of these
conditions are independent of the parameter $m$ which shows that
these conditions are valid for both the positive and inverse power
law potentials, however these conditions depend on the present value
of the field potential. Figure \textbf{2(a)} shows that the
condition $\rho^{eff}\geq0$, leads to $\beta<0$, which further
yields positive present value of the BD field potential $V_{0}$.
Moreover, the second condition will be satisfied if we take
$\beta<0$ with arbitrary $\omega$ and $\beta>0$ with $\omega>0$
only. Figure \textbf{2(b)} indicates that $\rho^{eff}+p^{eff}\geq0$
is satisfied for a particular choice of $\beta<0$ and
$-2<\omega<-1.5$. Clearly, the energy condition
$\rho^{eff}+p^{eff}\geq0$ is violated for $\beta>0$ with $\omega<0$.
It is easy to check that this choice of parameters
$(\beta=2,~\omega=-1.8)$ also leads to the violation of condition
imposed by scalar perturbations given by the expression
\begin{eqnarray}\nonumber
&&\frac{2H\phi^2+4\phi\dot{\phi}}{2H\phi+\dot{\phi}^2}-\frac{2\phi^2(2H\dot{\phi}+2\dot{H}\phi
+\ddot{\phi})}{(2H\phi+\dot{\phi}^2)^2}>a\phi,\\\nonumber
&&4\phi^2(\frac{X\omega/\phi-3H^2\phi-H\dot{\phi}/2}{(2H\phi+\dot{\phi}^2)^2})+3\phi>0
\end{eqnarray}
and consequently yields the ghost instabilities for the model.
However, the conditions imposed by tensor perturbation are trivially
satisfied.

\subsection{$f(R)$ Gravity}

\begin{figure} \centering
\epsfig{file=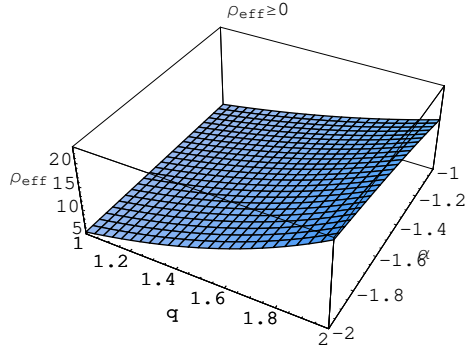,width=.45\linewidth} \caption{Plot shows WEC
versus free parameters $q$ and $\alpha$. Here we take
$q_0=-0.64,~H_0=0.718$ and $j_{0}=1.41$.}
\end{figure}
The action for $f(R)$ gravity described by
$S=\int\sqrt{-g}[\frac{M_{pl}^2}{2}f(R)+L_{m}]d^4x$ can be obtained
from the action (\ref{5}) for
\begin{eqnarray}\label{32}
K=-\frac{M_{pl}^2}{2}(Rf_{R}-f),\quad G_{3}=G_{5}=0,\quad
G_{4}=\frac{M_{pl}}{2}\phi,\quad \phi=M_{pl}f_{,R}.
\end{eqnarray}
Here $f(R)$ is an arbitrary function of the Rici scalar. Using these
values in energy conditions (\ref{14})-(\ref{17}), we obtain
\begin{eqnarray}\nonumber
\textbf{NEC}:\quad
&&(\ddot{R}-\dot{R}H)f''(R)+\dot{R}^2f'''(R)+\rho^m+p^m\geq0,
\end{eqnarray}
\begin{eqnarray}\nonumber
\textbf{WEC}:\quad&&\rho^{eff}+p^{eff}\geq0,
\quad\rho^m-\frac{1}{2}(f(R)-Rf'(R))-3H\dot{R}f''(R)\geq0,\\\nonumber
\textbf{SEC}:\quad&&\rho^{eff}+p^{eff}\geq0,\quad(f(R)-Rf'(R))+3(\ddot{R}+\dot{R}H)f''(R)\\\nonumber
&&+3\dot{R}^2f'''(R)+\rho^m+3p^m\geq0,\\\nonumber
\textbf{DEC}:\quad&&\rho^{eff}+p^{eff}\geq0,\quad\rho^{eff}\geq0,\quad-(5H\dot{R}+\ddot{R})f''(R)
-\dot{R}^2f'''(R)\\\nonumber
&&-(f(R)-Rf'(R))+\rho^m-p^m\geq0,
\end{eqnarray}
where prime denotes the derivative w.r.t. $R$. Let us consider the
example of logarithmic model in $f(R)$ gravity defined by
\cite{27}
\begin{eqnarray}
f(R)=R[\log(\alpha R)]^{q}-R;\quad\alpha<0,\quad q>0.
\end{eqnarray}
In this case, the WEC $(\rho^{eff}\geq0)$ leads to
\begin{eqnarray}\nonumber
&&-6H^2_{0}(1-q_{0})[\log(-6\alpha
H_{0}^2(1-q_{0}))]^q+6H_0^2(1-q_0)+6H_0^2(1-q_0)\\\nonumber
&&\times[(\log(-6\alpha H_{0}^2(1-q_{0})))^{q}+q(\log(-6\alpha
H_{0}^2(1-q_{0})))^{q-1}-1]-36H_{0}^4\\\nonumber
&&\times(j_0-q_0-2)[\frac{q}{6H_0^2(1-q_0)}(\log(-6\alpha
H_{0}^2(1-q_{0}))^{q})+\frac{q(q-1)}{6H_0^2(1-q_0)}\\\label{33}
&&\times(\log(-6\alpha H_{0}^2(1-q_{0})))^{q-2}]\geq0,
\end{eqnarray}
where we have used $\dot{R}=-6H^2(1-q)$ and the present values of
respective parameters. Notice that we have taken only the
condition $\rho^{eff}\geq0$ as the other condition involves the
present value of snap parameter $s$ which is not correctly
estimated in literature yet. Figure \textbf{3} shows that the WEC
is satisfied for a suitable range of both the parameters $q$ and
$\alpha$.

\subsection{Kinetic Gravity Braiding Model}

\begin{figure}
\centering \epsfig{file=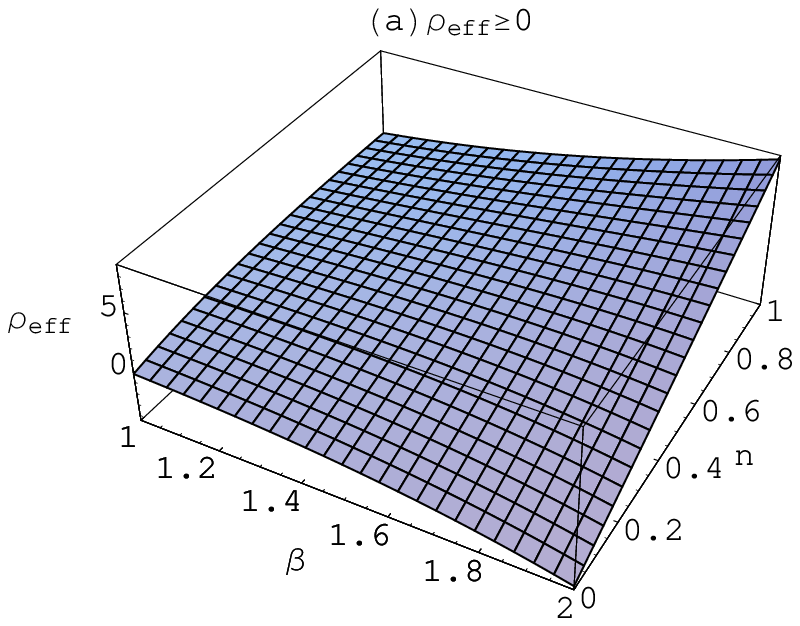,width=.45\linewidth}
\epsfig{file=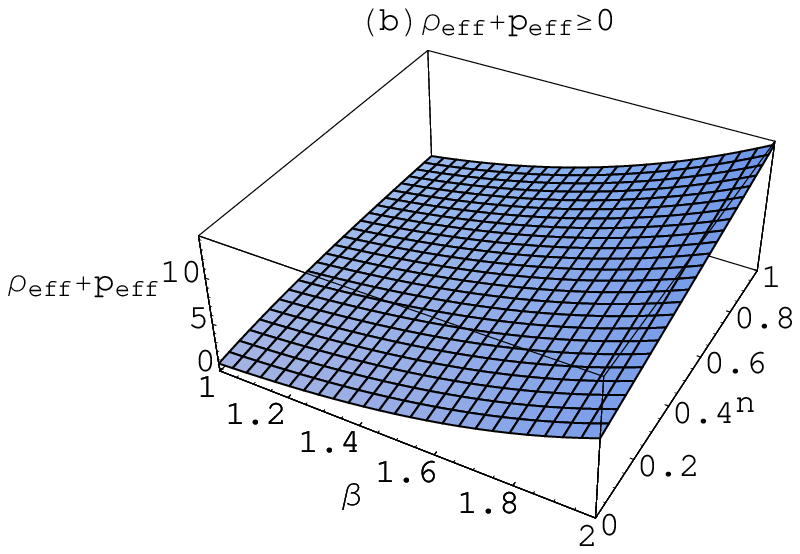,width=.45\linewidth} \caption{Plots (a) and
(b) show WEC namely $\rho^{eff}\geq0$ and $\rho^{eff}+p^{eff}\geq0$
versus parameters $n$ and $\beta$.}
\end{figure}
Kinetic gravity braiding is defined by the action \cite{28}
\begin{eqnarray}\nonumber
S&=&\int
d^4x\sqrt{-g}(K(\phi,X)-G_{3}(\phi,X)\Box\phi+\frac{M_{pl}^2}{2}R
+\frac{M_{pl}^2}{2}[(\Box\phi)^2\\\nonumber
&-&(\nabla_\mu\nabla_{\nu}\phi)(\nabla^\mu\nabla^\nu\phi)]+L_{m}),
\end{eqnarray}
where the functions $K$ and $G_{3}$ are arbitrary (while other
functions are taken to be zero in action (\ref{5}). A particular
choice of these functions proposed by Dvali and Turner \cite{29}
is given by $K=-X$ and $G_{3}=cX^n$, where $c$ and $n$ are
constants. For this choice of model, the respective energy
conditions turn out to be
\begin{eqnarray}\nonumber
\textbf{NEC}:\quad&&
\frac{1}{M_{pl}^2}[-2X+6ncH\dot{\phi}X^n-2cnX^n\ddot{\phi}+\rho^m+p^m]\geq0,\\\nonumber
\textbf{WEC}:\quad&&\rho^{eff}+p^{eff}\geq0,
\quad\frac{1}{M_{pl}^2}[-X+6ncH\dot{\phi}X^n+\rho^m]\geq0,\\\nonumber
\textbf{SEC}:\quad&&\rho^{eff}+p^{eff}\geq0,
\quad\frac{1}{M_{pl}^2}[-4X+6ncH\dot{\phi}X^n-6cnX^n\ddot{\phi}\\\nonumber
&&+\rho^m+3p^m]\geq0,\\\nonumber
\textbf{DEC}:\quad&&\rho^{eff}+p^{eff}\geq0,\quad\rho^{eff}\geq0,
\quad\frac{1}{M_{pl}^2}[6ncH\dot{\phi}X^n+2cnX^n\ddot{\phi}\\\nonumber
&&+\rho^m-p^m]\geq0.
\end{eqnarray}
By taking the power law evolution of the scalar field, WEC
yields
\begin{eqnarray}\label{34}
&&[-\frac{\beta^2a^{2\beta}H^2}{2}+6ncH\dot{\phi}
(\frac{\beta^2a^{2\beta}H^2}{2})^n]\geq0,\\\label{35}
&&[-2\frac{\beta^2a^{2\beta}H^2}{2}+6ncH\dot{\phi}
(\frac{\beta^2a^{2\beta}H^2}{2})^n-2cn(\frac{\beta^2a^{2\beta}H^2}{2})^n\ddot{\phi}]\geq0.
\end{eqnarray}
Here we have taken the present values of the ordinary density and
pressure to be zero. Using the present values of the Hubble
parameter and the scale factor, WEC can be satisfied only when both
the parameters $n$ and $\beta$ remain positive as indicated in
Figure \textbf{4}, where we have taken $0<n<1$ and $1<\beta<2$. In
this case, scalar perturbations lead to the constraints
\begin{eqnarray}\nonumber
&&H^2(2+q)-\frac{cnHX^n\dot{\phi}}{a}+cn^2X^{n-1}\dot{X}\dot{\phi}+cnX^n\ddot{\phi}>(H-cnX^n\dot{\phi})^2,\\\nonumber
&&(\frac{-X+12cnHX^n\dot{\phi}+6nc(n-1)HX^n\dot{\phi}-3H^2M_{pl}^2}{(HM_{pl}^2-cnX^n\dot{\phi})^2})M_{pl}^2>-3.
\end{eqnarray}
It is easy to check that the energy conditions are violated for
$0.6<n<1$ and negative range of $\beta$ (e.g., $\beta=-10$). For
this choice of parameters, the above constraints are also violated
and hence the ghost instabilities occur. However, constraints
imposed by tensor perturbations are trivially satisfied as
$M_{pl}^2\phi>0$.

\subsection{Covariant Galilean Model}

In the absence of potential, the covariant Galilean model
\cite{29} is defined by the following choice of parameters in
action (\ref{5})
\begin{eqnarray}\label{36}
K=-c_{2}X,\quad G_{3}=\frac{c_{3}}{M^3}X,\quad
G_{4}=\frac{M_{pl}^2}{2}-\frac{c_{4}}{M^6}X^2, \quad
G_{5}=\frac{3c_{5}}{M^9}X^2,
\end{eqnarray}
where $c_{2},~c_{3},~c_{4}$ and $c_{5}$ are dimensionless
constants while $M$ is constant with dimensions of mass. Using
these values in Eqs.(\ref{14})-(\ref{17}), it follows that
\begin{eqnarray}\nonumber
&&\textbf{NEC}:\quad\frac{1}{2(\frac{M_{pl}^2}{2}-\frac{c_{4}}{M^6}X^2)}
[-2Xc_2+6XH\dot{\phi}\frac{c_3}{M^3}-72H^2X^2\frac{c_4}{M^6}
\\\nonumber
&&+60H^3X^2\frac{c_5}{M^9}-2X\ddot{\phi}\frac{c_3}{M^3}+24HX\dot{X}\frac{c_4}{M^6}
+16\dot{H}X^2\frac{c_4}{M^6}-12X^2\\\nonumber&&
\times(2H\dot{H}\dot{\phi}+3H^2\ddot{\phi})\frac{c_5}{M^9}-24H^2X^2\ddot{\phi}\frac{c_5}{M^9}+\rho^m+p^m]\geq0,\nonumber
\end{eqnarray}
\begin{eqnarray}\nonumber
&&\textbf{WEC}:\quad\rho^{eff}+p^{eff}\geq0,\quad\frac{1}{2(\frac{M_{pl}^2}{2}-\frac{c_{4}}{M^6}X^2)}[-Xc_2
+6XH\dot{\phi}\frac{c_3}{M^3}\\\nonumber
&&-96H^2X^2\frac{c_4}{M^6}+84H^3X^2\dot{\phi}\frac{c_5}{M^9}+\rho^m]\geq0,\\\nonumber
&&\textbf{SEC}:\quad\rho^{eff}+p^{eff}\geq0,\quad\frac{1}{2(\frac{M_{pl}^2}{2}
-\frac{c_{4}}{M^6}X^2)}[-4Xc_2+6XH\dot{\phi}\frac{c_3}{M^3}-6X\ddot{\phi}\\\nonumber&&\times
\frac{c_3}{M^3}+24XH^2\frac{c_4}{M^6}-48H^2X^2\frac{c_4}{M^6}+12H^3X^2\dot{\phi}\frac{c_5}{M^9}
-180H^2X^2\ddot{\phi}\frac{c_5}{M^9}\\\nonumber
&&-72X^2H\dot{H}\dot{\phi}\frac{c_5}{M^9}+48X\dot{X}H\frac{c_4}{M^6}
+48X^2\dot{H}\frac{c_4}{M^6}+24HX\dot{X}\frac{c_4}{M^6}+72H^2X^2\\\nonumber
&&\times\frac{c_4}{M^6}+\rho^m+3p^m]\geq0,\\\nonumber
&&\textbf{DEC}:\quad\rho^{eff}+p^{eff}\geq0, \quad\rho^{eff}\geq0,
\quad\frac{1}{2(\frac{M_{pl}^2}{2}-\frac{c_{4}}{M^6}X^2)}
[6XH\dot{\phi}\frac{c_3}{M^3}\\\nonumber
&&+2X\ddot{\phi}\frac{c_3}{M^3}-72H^2X^2\frac{c_4}{M^6}
-48XH^2\frac{c_4}{M^6}+108H^3X^2\dot{\phi}\frac{c_5}{M^6}-24HX\dot{X}\frac{c_4}{M^6}
\\\nonumber&&
-16\dot{H}X^2\frac{c_4}{M^6}+12X^2(2H\dot{H}\dot{\phi}+2H^2\ddot{\phi})\frac{c_5}{M^9}
+24H^2X^2\ddot{\phi}\frac{c_5}{M^9}+\rho^m-p^m]\geq0.
\end{eqnarray}

By taking the power law anstaz for scalar field and consequently,
for kinetic term $X$, the WEC in terms of present values of the
involved parameter require the following inequalities
\begin{eqnarray}\nonumber
&&\frac{1}{(1-\frac{c_{4}\beta^4H_{0}^4}{2M^6})}
[-c_2H_{0}^2\beta^2+\frac{3H_{0}^4c_3\beta^3}{M^3}-\frac{18H_{0}^6\beta^4c_4}{M^6}
+\frac{15H_0^7\beta c_5}{M^9}\\\nonumber
&&-\beta^3H_0^4(\beta-1-q_0)\frac{c_3}{M^3}+24H_{0}^6\beta^4(\beta-1-q_0)
\frac{c_4}{M^6}-16H_0^6\beta^4\\\nonumber
&&\times(1+q)\frac{c_4}{M^6}-3H_0^4\beta^4(3H_0^4\beta(\beta-1-q_0)
-2H_0^4\beta(1+q))\frac{c_5}{M^9}\\\label{37}
&&-6H_0^8\beta^5(\beta-1-q_0)\frac{c_5}{M^9}]\geq0,\\\label{38}
&&\frac{1}{(1-\frac{c_{4}\beta^4H_{0}^4}{2M^6})}
[-\frac{H_{0}^2\beta^2c_2}{2}+\frac{3c_3H_{0}^4\beta^3}{M^3}
-\frac{24H_{0}^6\beta^4c_4}{M^6}+\rho_{0}^m]\geq0.
\end{eqnarray}
Clearly, these conditions are satisfied when both $G_4$ and terms
inside the brackets are positive. Since $G_4>0$ requires
$\beta>(\frac{7.5M^6}{c_4})^{1/4}$, therefore a suitable choice of
all these parameters yield the consistency with WEC if parameter
$\beta$ remains small and positive $(\beta<7.8)$ while $c_2$ remains
negative as shown in Figure \textbf{5}. Here we have taken
$-5<c_2<-1$ or $-50<c_2<-10$ and $5.5<\beta<8$.
\begin{figure}
\centering \epsfig{file=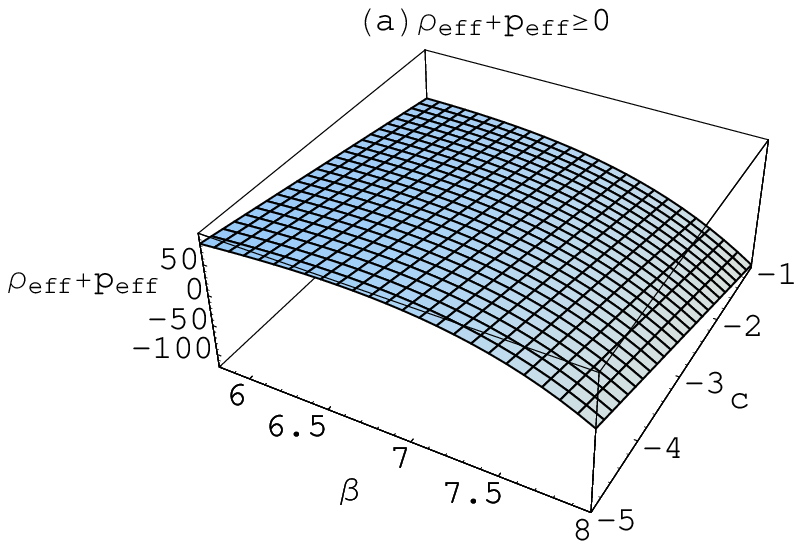,width=.45\linewidth}
\epsfig{file=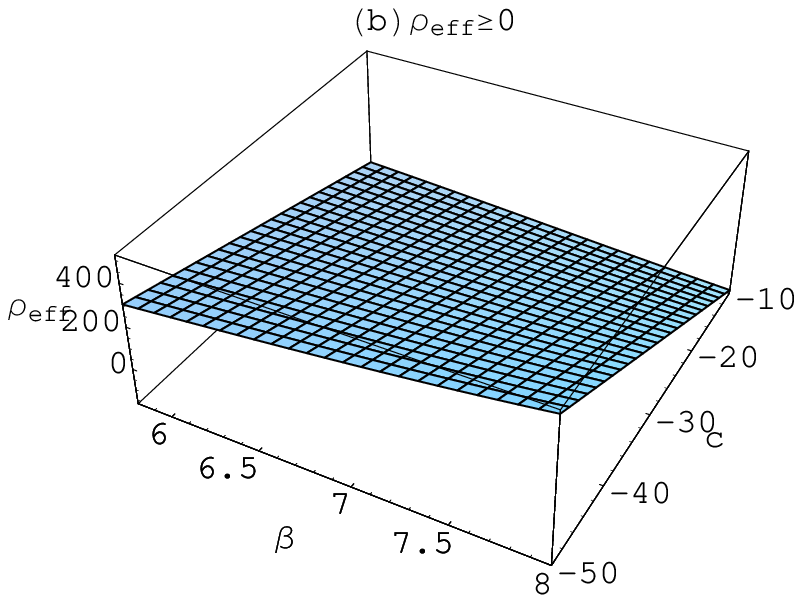,width=.45\linewidth} \caption{Plots (a) and
(b) show WEC versus parameters $c_2$ and $\beta$ with
$c_3=2,~c_4=4,~c_5=3$ and $M=2$.}
\end{figure}
In this case, tensor perturbations suggest the following conditions
for the avoidance of ghost instabilities
\begin{eqnarray}\nonumber
\mathcal{F}_T&=&M_{pl}^2-2X^2\frac{c_4}{M^6}-12X^2\ddot{\phi}\frac{c_5}{M^9}>0,\\\nonumber
\mathcal{G}_T&=&M_{pl}^2+6X^2\frac{c_4}{M^6}-12X^2\dot{\phi}H\frac{c_5}{M^9}>0.
\end{eqnarray}
Clearly, the energy conditions are violated for $\beta>8$ with
$-5<c_2<-1$. For this choice of parameters, the above constraints
imposed by tensor perturbations are also violated and consequently,
the ghost instabilities exist. However, scalar perturbations lead to
very complicated expressions, so we consider only the conditions
imposed by tensor perturbations.

\section{Summary}

The most general scalar-tensor theory being a combination of
various DE proposals provides a vast gravitational framework for
the discussion of accelerated expansion of the universe. The
modified theories involve some extra degrees of freedom that are
described by the models with some unknown parameters. It would be
interesting to restrict these parameters on physical grounds. In
cosmology, this can be done by making compatibility with local
gravity tests. In a gravitational theory, energy conditions can be
used as an approach to restrict these parameters. In the present
paper, we consider the most general scalar tensor gravity with the
field equations involving second-order derivatives. Firstly, we
have explored the effective energy-momentum tensor and its trace
by inverting the generalized field equations which can be used to
find the energy condition bounds for any spacetime manifold.

In order to describe these conditions for specific cases, we
consider flat FRW universe model with perfect fluid. By defining the
effective energy density and pressure, we have expressed the strong,
weak, null and dominant energy conditions. For the sake of
convenience, we have assumed that the scalar field evolves as a
power of scale factor. Also, the derivative terms are removed by
expressing these conditions in terms of cosmological quantities like
deceleration, snap and jerk parameters. An estimation to the present
values of these parameters is available in literature \cite{24} that
can be used to find the constraints on free parameters of the model.
The derived energy conditions are the most general in nature
involving many arbitrary functions $K,~G_3, G_4$ and $G_5$ that
correspond to different DE proposals.

For the application of these energy conditions, we have taken
different choices of the functions $K,~G_3,~G_4$ and $G_5$ and
have deduced the energy conditions for k-essence model, BD
gravity, $f(R)$ theory, kinetic gravity braiding and covariant
Galilean model. In each case, there are many free parameters. It
is not possible to find their range that would be consistent with
energy conditions as well. For this reason, we have specified some
of these parameters and restricted the others. The results can be
summarized as:
\begin{itemize}
\item The WEC in k-essence model is satisfied only if
the free parameters satisfy $0<M_0<1,~\beta>1.4$ with arbitrary
$B$ and $\beta<1.4$ with $B>10$.
\item In literature \cite{20}, using the equivalence between BD and $f(R)$
gravity $(\omega=0)$, it has been shown that $V_0$ should be
negative. In our case ($\omega\neq0$), the WEC restricts the
parameter $\beta$ to be negative (the scalar field should be of
contractive nature) for the positive present value of BD field
potential, i.e., $V_{0}>0$ which is physically correct. However, the
condition $\rho^{eff}+p^{eff}\geq0$ is satisfied only when we take
$\beta<0$ with arbitrary $\omega$ and $\beta>0$ with $\omega>0$.
Thus it can be concluded that expanding scalar field with negative
range of BD parameter allowed for cosmic expansion is inconsistent
with the WEC.
\item In $f(R)$ gravity, many authors have used the energy conditions to
find the constraints on the models like $f(R)=\alpha R^n$ or
$R+\alpha R^n$ \cite{20}. In the present case, we have found the
constraints on the logarithmic $f(R)$ model. It is seen that the
restrictions on free parameters $\alpha<0$ and $q>0$ are consistent
with WEC.
\item In kinetic gravity braiding, the WEC is
satisfied for the presented model only when $\beta>0$ (that shows
expanding scalar field) and $n>0$.
\item In covariant Galilean model of DE, the WEC is satisfied when
free parameters of the model satisfy
$\beta>(\frac{7.5M^6}{c_4})^{1/4}$ and $c_2<0$.
\end{itemize}
All these results are also shown through graphs. Further, we have
determined the conditions for the avoidance of ghost instabilities
using the constraints based on the scalar and tensor perturbations
proposed by Kobayashi et al. \cite{14}. It is concluded that the
violation of these energy conditions leads to the occurrence of
ghost and gradient instabilities in the above mentioned cases of the
most general second-order scalar tensor theory. It would be
interesting to investigate the constraints on other DE models like
exponential model of $f(R)$ gravity and other forms of potentials
for BD gravity etc. by making them consistent with the energy
conditions.

\vspace{0.3cm}

\renewcommand{\theequation}{A\arabic{equation}}
\setcounter{equation}{0}
\section*{Appendix A}
\begin{eqnarray}\nonumber
T^{(\phi)}_{\mu\nu}&=&\frac{1}{2}K_{X}\nabla_\mu\phi\nabla_\nu\phi+\frac{1}{2}Kg_{\mu\nu}
-\frac{1}{2}G_{3X}\Box\phi\nabla_{\nu}\phi\nabla_{\mu}\phi-\nabla_{(\mu}G_3\nabla_{\nu)}\phi
\\\nonumber
&+&\frac{1}{2}g_{\mu\nu}\nabla_{\lambda}G_{3}\nabla^{\lambda}\phi
+\frac{1}{2}G_{4X}R\nabla_{\mu}\phi\nabla_{\nu}\phi+\frac{1}{2}G_{4XX}[(\Box\phi)^2
-(\nabla_{\alpha}\nabla_{\beta}\phi)^2]\\\nonumber
&\times&\nabla_{\mu}\phi\nabla_{\nu}\phi+G_{4X}\Box\phi\nabla_{\mu}\nabla_{\nu}\phi
-G_{4X}\nabla_{\lambda}\nabla_\mu\phi\nabla^\lambda\nabla_\nu\phi-2\nabla_\lambda
G_{4X}\\\nonumber
&\times&\nabla^\lambda\nabla_{(\mu}\phi\nabla_{\nu)}\phi+\nabla_{\lambda}G_{4X}
\nabla^\lambda\phi\nabla_\mu\nabla\nu\phi-g_{\mu\nu}(G_{4\phi}\Box\phi-2XG_{4\phi\phi})\\\nonumber
&-&g_{\mu\nu}[-2G_{4X\phi}\nabla_\alpha\nabla_\beta\phi\nabla^\alpha\phi\nabla^\beta\phi
+G_{4XX}\nabla_\alpha\nabla_\lambda\phi\nabla_\beta\nabla^\lambda\phi\nabla^\alpha\phi\nabla^\beta\phi\\\nonumber
&+&\frac{1}{2}G_{4X}[(\Box\phi)^2-(\nabla_\alpha\nabla_\beta\phi)^2]]
-2[G_{4X}R_{\lambda(\mu}\nabla_{\nu)}\phi\nabla^\lambda\phi-\nabla_{(\mu}G_{4X}\nabla_{\nu)}\\\nonumber
&\times&\phi\Box\phi]+g_{\mu\nu}[G_{4X}R^{\alpha\beta}\nabla_{\alpha}\phi\nabla_\beta\phi-\nabla_\lambda
G_{4X}\nabla^\lambda\phi\Box\phi]-G_{4X}R_{\mu\alpha\beta\nu\beta}\\\nonumber
&\times&\nabla^\alpha\phi\nabla^\beta\phi+G_{4\phi\nabla_\mu\nabla_\nu\phi}+G_{4\phi\phi}\nabla_\mu\phi\nabla_\nu\phi
-2G_{4X\phi}\nabla^\lambda\phi\nabla_\lambda\nabla_{(\mu}\phi\nabla_{\nu)}\phi\\\nonumber
&+&G_{4XX}\nabla^\alpha\phi\nabla_\alpha\nabla_\mu\phi\nabla^\beta\phi\nabla_\beta\nabla_\nu\phi
-G_{5X}R_{\alpha\beta}\nabla^\alpha\phi\nabla^\beta\nabla_{(\mu}\phi\nabla_{\nu)}\phi
+G_{5X}\nonumber
\end{eqnarray}
\begin{eqnarray}\nonumber
&\times&R_{\alpha(\mu}\nabla_{\nu)}\phi\nabla^\alpha\phi\Box\phi+\frac{1}{2}G_{5X}
R_{\alpha\beta}\nabla^\alpha\phi\nabla^\beta\phi\nabla_\mu\nabla_\nu\phi
\frac{1}{2}G_{5X}R_{\mu\nu\alpha\beta}\nabla^\alpha\phi\\\nonumber
&\times&\nabla^\beta\phi\Box\phi-G_{5X}R_{\alpha\lambda\beta(\mu}\nabla_{\nu)}\phi\nabla^\lambda\phi
\nabla^\alpha\nabla^\beta\phi-G_{5X}R_{\alpha\lambda\beta(\mu}\nabla_{\nu)}\nabla^\lambda\phi
\nabla^\alpha\phi\\\nonumber
&\times&\nabla^\beta\phi+\frac{1}{2}\nabla_{(\mu}[G_{5X}\nabla^\alpha\phi]\nabla_\alpha\nabla_{\nu)}\phi\Box\phi
-\frac{1}{2}\nabla_{(\mu}[G_{5\phi}\nabla_{\nu)}\phi]\Box\phi\\\nonumber
&+&\nabla_{\lambda}[G_{5\phi}\nabla_{(\mu}\phi]\nabla_{\nu)}\nabla^\lambda\phi-\frac{1}{2}
[\nabla_\lambda(G_{5\phi}\nabla^\lambda\phi)-\nabla_\alpha(G_{5X}\nabla_\beta\phi)\nabla^\alpha\nabla^\beta\phi]
\\\nonumber &\times&\nabla_\mu\nabla_\nu\phi-\nabla^\alpha
G_{5}\nabla^\beta\phi
R_{\alpha(\mu\nu)\beta}+\nabla_{(\mu}G_{5}G_{\nu)\lambda}\nabla^\lambda\phi
-\frac{1}{2}\nabla_{(\mu}G_{5X}\nabla_{\nu)}\phi\\\nonumber
&\times&[(\Box\phi)^2-(\nabla_{\alpha}\nabla_{\beta}\phi)^2]+\nabla^\lambda
G_{5}R_{\lambda(\mu}\nabla_{\nu)}\phi-\nabla_\alpha[G_{5X}\nabla_\beta\phi]
\nabla^\alpha\nabla_{(\mu}\phi\\\nonumber
&\times&\nabla^{\beta}\nabla_{\nu)}\phi+\nabla_\beta
G_{5X}[\Box\phi\nabla^\beta\nabla_\mu\phi-\nabla^\alpha\nabla^\beta\phi\nabla_\alpha\nabla_{(\mu\phi}]\nabla_{\nu)}\phi
-\frac{1}{2}\nabla^\alpha\phi\\\nonumber
&\times&\nabla_{\alpha}G_{5X}[\Box\phi\nabla_{\mu}\nabla_\nu\phi-\nabla_{\beta}\nabla_\mu
\phi\nabla_{\beta}\nabla_\nu\phi]+\frac{1}{2}G_{5X}G_{\alpha\beta}\nabla^\alpha
\nabla^\beta\phi\nabla_\mu\phi\nabla_\nu\phi\\\nonumber
&+&\frac{1}{2}G_{5X}\Box\phi\nabla_\alpha\nabla_\mu\phi\nabla^\alpha\nabla_\nu\phi-\frac{1}{2}
G_{5X}(\Box\phi)^2\nabla_\mu\nabla_\nu\phi-\frac{1}{12}G_{5XX}[(\Box\phi)^3\\\nonumber
&-&3(\Box\phi)(\nabla_\alpha\nabla_\beta\phi)^2+(\nabla_\alpha\nabla_\beta\phi)^3]\nabla_\mu\phi\nabla_\nu\phi
-\frac{1}{2}\nabla_\lambda
G_{5}G_{\mu\nu}\nabla^\lambda\phi-g_{\mu\nu}\\\nonumber
&\times&\{-\frac{1}{6}G_{5X}[(\Box\phi)^3-3(\Box\phi)(\nabla_\alpha\nabla_\beta\phi)^2
+(\nabla_\alpha\nabla_\beta\phi)^3]\nabla_\alpha
G_{5}R^{\alpha\beta}\nabla_\beta\phi \\\nonumber
&-&\frac{1}{2}\nabla_\alpha(G_{5\phi}\nabla^\lambda\phi)\Box\phi+\frac{1}{2}\nabla_\alpha
G_{5\phi}\nabla_\beta\phi\nabla^\alpha\nabla^\beta\phi-\frac{1}{2}\nabla_\alpha
G_{5X}\nabla^\alpha X\Box\phi\\\nonumber &+&\frac{1}{2}\nabla_\alpha
G_{5X}\nabla_\beta
X\nabla^\alpha\nabla^\beta\phi-\frac{1}{4}\nabla^\lambda
G_{5X}\nabla_\lambda\phi[(\Box\phi)^2
-(\nabla_{\alpha}\nabla_{\beta}\phi)^2]\\\label{100}
&+&\frac{1}{2}G_{5X}R_{\alpha\beta}\nabla^\alpha\phi\nabla^\beta\phi\Box\phi
-\frac{1}{2}G_{5X}R_{\alpha\lambda\beta\rho}\}.
\end{eqnarray}

\end{document}